# Amplified Spontaneous Emission enhanced Forward Stimulated Raman Scattering in dye solutions

J.A. Dharmadhikari [a], A.K. Dharmadhikari [a], Alpana. Mishra [b] and G. Ravindra Kumar. [a]

a) Tata Institute of Fundamental Research, 1 Homi Bhabha Road Colaba, Mumbai, India 400005. Fax: +91-22-2152110 E-mail: grk@tifr.res.in
b) Department of Chemistry and Chemical Engineering Polytechnic University, 6 Metrotech Center, Brooklyn, NY USA 11201

**Abstract**

We study forward stimulated Raman emission from weakly fluorescent dye 4'-diethylamino-*N*-methyl-4-stilbazolium tosylate (DEST) in 1,2,dichloroethane solution excited by a 28 ps, 532 nm Nd: YAG laser. Neat 1, 2, dichloroethane emits the first Stokes line at 631 nm with a spectral width of 1.6 nm corresponding to a Raman shift of 2956 $cm^{-1}$. We observe reduction of spectral width with the addition of DEST in 1, 2, dichloroethane solution. The single pass conversion efficiency for forward Raman emission is as high as 20% in a 1 cm path length sample. The pulse duration of forward stimulated Raman emission measured by a third order autocorrelation technique is 10 ps in neat 1, 2, dichloroethane, whereas it is ~3 ps for 4 x $10^{-5}$ mol/l of DEST solution.

# 1   Introduction

Stimulated Raman scattering (SRS) has been well investigated in a variety of materials [1]. It offers a simple and inexpensive way of tuning the wavelength of a laser system. The stimulated Raman process originates from spontaneous Raman scattering and requires high pump intensities, reducing the conversion efficiency of the process. Further, at sufficiently high intensities, other non-linear optical effects such as stimulated Brillouin scattering start appearing. In ultrashort pulse interactions, SRS has to compete with continuum generation [2, 3]. Methods such as external seeding, either by way of injection of radiation into the Raman medium at a Stokes shifted wavelength [4], or by addition of lasing dye particularly for liquids that fluoresce at the Raman lines of the medium [5] are used to reduce the threshold of stimulated Raman process. The addition of lasing dye into a neat solvent leads to interesting competition between SRS from the solvent and amplified spontaneous emission (ASE) from the dye, provided there is spectral overlap between the two processes. Under such conditions, SRS, fluorescence and ASE cannot be treated independently and SRS builds up mostly from the dye fluorescence. In addition, if there is sufficient stimulated emission from the dye, the Raman gain can be boosted substantially. Another important feature of SRS is the shortening of the Stokes pulse width from that of the input laser pulse. Pulse compression factors as high as 40 have been observed [6] in backward SRS [7-12]. Such a large compression is possible due to the fact that backscattered pulse continuously encounters an undepleted pump light giving rise to high amplification and sharpening. On the other hand a forward traveling Stokes pulse has access only to the pump energy stored in the traveling volume region occupied. In addition, saturation due to pump depletion limits the forward Stokes emission. However, Forward SRS has some intrinsic advantages over its backward counterpart in terms of simplicity of set up, ease of design of the optics, separation from pump beam etc.  There is, therefore, a pressing need to find methods that can enhance forward SRS.  The addition of a dye whose fluorescence acts as a seed for SRS is one such option.  The availability of population inversion in the dye can further enhance the forward Raman gain dramatically.

Intense SRS, though in the backward geometry, has been reported previously [5] using highly fluorescent dyes such as Rhodamine-6G and DCM. Recently, one of the authors has reported highly efficient mirrorless lasing using single and multi-photon absorption in weakly fluorescent stibazolium salt styryl pyridinium cyanine dye (SPCD) or 4′-dimethylamino-*N*-methyl-4-stilbazolium methylsulfate (DMSM) [13,14]. These belong to a new class of molecules that are weak in florescence but surprisingly have ASE. The high ASE efficiency (40%) has been attributed to large dipole moments associated with the charge transfer transition in the organic molecular salts. It is therefore interesting to explore if the strong ASE from these molecules can enhance SRS (particularly the forward process) efficiencies.

In this paper we investigate forward SRS from solutions of 4'-diethylamino-*N*-methyl-4-stilbazolium tosylate (DEST) in 1, 2 dichloroethane. We exploit the strong spectral overlap of the first Stokes line of 1, 2 dichloroethane with the ASE of the dye in this solvent. Unlike some previous observations [4, 8], we achieve the seeding and amplification of SRS in a single stage. We also report the spectral evolution of the forward Raman emission at various pump intensities. We demonstrate a three-fold increase in efficiency in comparison to the neat solvent. We achieve a pulse compression factor of 3 in the dye solution as compared to that from the neat solvent, giving an overall compression factor of ~10 from the pump pulse duration.

## 2. Experimental
### 2.1 Experimental set up

The experiment uses an Nd:YAG laser (Continuum PY61C-10) emitting 28 ps, 532 nm pulses at 10 Hz as the pump source. The laser beam is focused with a lens of focal length 30 cm into a 1 cm path cuvette, containing neat 1, 2, dichloroethane and the generated first stokes Raman signal is detected in the forward direction by using a single shot spectrometer (Ophir 'Wavestar V'). In the same setup DEST is added to the neat solvent to detect the generated ASE and Raman signals from the solution.

The pulse duration of the Raman emission is measured by third-order auto correlation, employing two beam self-diffraction in a thin gel film doped with brilliant green dye (Fig.1). The Raman beam is split in two parts with a beam splitter and the beams are spatially overlapped in the dye film. The angle between the two beams is < 2 $^o$ and the beams are focused on the film by a lens of focal length 30cm. The first order diffracted signal arises from third order nonlinearity, and is sensitive to the temporal overlap between the beams. The variation of this signal with respect to the temporal separation between the beams is, therefore, a measure of the pulse duration [15, 16]. The first order diffracted signal is detected by a photodiode PD2; and the fluctuations in the laser energy are monitored by photodiode PD1. The photodiodes signals are acquired using a digital storage oscilloscope (Yokogawa DL 7200). Each data point given here corresponds to an average of 50 shots.

## 2.2. Results

The linear absorption spectrum of DEST at low concentration ($10^{-6}$ mol/l) in methanol shows that the absorption peak is located at 490 nm. Further, fluorescence quantum efficiency is ~0.5% for DEST, which is much lower than that for the well-known laser dye Rhodamine6G (R6G). The emission spectra show ASE centered at 617 nm with a spectral width of 20 nm [17].

SRS from neat 1, 2, dichloroethane has first and second Stokes emission at 631 nm and 776 nm as well as first anti-Stokes line at 459nm. Fig.2 shows only the first Stokes signal at 631 nm having a line width of 1.6 nm at the input energy of 1.1 mJ (peak focused intensity of 800 GWcm$^{-2}$). Note that this line has considerable overlap with the emission spectrum of DEST solution.Fig.3 shows the emission spectrum of DEST in 1, 2, dichloroethane at different pump energies. The spectra show only ASE with a spectral width of 20 nm at 0.3 mJ (Fig. 3a). At the input energy of 0.7 mJ an additional peak centered at 631 nm due to SRS from the solvent along with the ASE starts appearing (Fig. 3b). The second Stokes and first anti-Stokes lines observed in the neat solvent are suppressed with the addition of dye. At a higher input energy of 1.1 mJ, the Raman signal amplitude rapidly increases as compared to the ASE signal indicating the dominance of

Raman process (Fig. 3c). The ASE peak significantly reduces at 1.3 mJ and the spectrum has a dominance of the SRS signal. Moreover, the line width of the generated Raman signal (0.8nm) is reduced compared to that from 1, 2, dichloroethane (1.6nm). We measured the line width of a He-Ne laser (which is close to the observed Raman line) with the spectrometer to be 0.8 nm, which is much higher than the true line width of the laser. Thus, we are limited by the instrument resolution and the actual Raman line width may be smaller

Fig.4 gives the variation of the Raman signal with the input laser energy. Above a pump energy of 0.9 mJ, Raman emission increases rapidly in the case of dye solution as compared to that from neat solvent. The energy of the first Stokes line is measured to be 0.1 mJ at a pump input of 1 mJ giving a conversion efficiency of ~20% (with pump corrected for linear absorption in the solution) for $4 \times 10^{-5}$ mol/l DEST as compared to 8% in neat solvent. This enhancement is clearly due to the overlap between the ASE spectra of the dye and Raman emission of 1, 2, dicholroethane. At this concentration, SRS essentially builds up from the ASE of dye, which is stronger than the spontaneous Raman noise, leading to enhancement in the SRS efficiency. Thus with the addition of small quantity of dye molecules in the solvent the efficiency is enhanced. At higher concentration of $7 \times 10^{-4}$ mol/l there is a drastic reduction in efficiency, making it difficult to quantify. This is expected since at higher concentration the dye (particularly that close to the front surface of the cuvette) absorbs most of the pump energy producing large ASE, thereby reducing the energy available for Raman generation. This indicates that there is an optimum concentration for the ASE seeding of SRS.

We can also view this as competition between two different processes, (1) a coherent process (SRS), which has a higher threshold and (2) and incoherent process (ASE) which has a lower threshold. At low laser energies ~0.3 mJ, the incoherent process dominates (fig. 3a) At this intensity, there is a small hump seen in the spectrum at 631 nm due to the Raman shifted emission from the solvent. As the laser energy increases two distinct peaks appear in the spectra indicating the presence of both the processes-one broad peak centered at 620 nm due to the ASE from the dye and a narrow peak centered at 631 nm because of the stimulated Raman scattering from the solvent. Thus at these pumping energies ASE provides sufficient seeding for the stimulated Raman process. Moreover, the line width is reduced (to instrument limit of 0.8 nm) as compared to 1.6 nm in neat 1, 2 dichloroethane at 1.3mJ energy. This can be seen as an interesting conversion of incoherent light to a coherent laser beam and offers great promise for the design of efficient, tunable coherent light sources in widely different wavelength regions. The reduction in the line width on addition of dye molecules in the solvent clearly indicates that the gain is enhanced for the first stokes line. This can be explained by considering the steady state pumping conditions, $\tau \gg T_2 = 1/\pi\Delta\nu_R$, where $\tau$ is the pump pulse duration, $T_2$ is the dephasing (relaxation) time of the SRS-active vibration mode of solvent, and $\Delta\nu_R$ is the line width of the corresponding Raman shifted line in the spontaneous Raman scattering spectrum. If the intensity of pump (fundamental) laser radiation is much higher than the intensity of the first Stokes emission $I_f \gg I_{St1}$, the SRS amplification at the first Stokes emission can be written as [1],

$$\frac{dI_{St1}(l)}{dl} = g_R I_f(l) I_{St1}(l) + I_f(l)\frac{d\mathbf{s}}{d\Omega} N\Delta\Omega$$

$g_R = \dfrac{2 I_{St1}^2 N}{p n_{st1}^2 h \mathbf{n}_f \Delta \mathbf{n}_R} \dfrac{d\mathbf{s}}{d\Omega}$ is the steady-state Raman gain coefficient, $\dfrac{d\mathbf{s}}{d\Omega}$ is the differential cross-section per molecule, $\Delta\Omega$ is the solid angle and N is the number density. This is in agreement with the above results, as the line width $\Delta\nu_R$ is inversely related to the gain coefficient.

We now discuss the effect of the DEST dye on the pulse duration SRS. Fig. 5a shows the trace of the pulse duration from a neat 1, 2, dichloroethane solution, with a FWHM ~10 ps. Thus there is nearly three times pulse compression as compared to the pump pulse duration of 28 ps. With addition of a small quantity of the dye (4 x $10^{-5}$ mol/l) the pulse duration further reduces to 3 ps as shown in Fig. 5b. This is again more than three-fold enhancement over the neat solvent and an overall ten-fold compression with respect to the pump pulse duration. Assuming a Gaussian laser pulse shape the compressed Stokes pulse duration ô$_s$ is given by [18],

ô$_s$ = ô$_p$/(ln (G$_{ss}$)$^{1/2}$

where $δ_p$ is the pump pulse duration and $G_{ss}$ is the steady state Raman gain of the medium and is given by $G_{ss} = g \times l$, where g is gain coefficient and l is the length of the gain medium. The gain coefficient is calculated from the above equation for neat solvent to be 3.2 cm/GW at input intensity of 800 GW/cm$^2$. This value agrees with the gain coefficients reported for other similar solvents [19].

## 3     Conclusions

In conclusion we report observations of aided and unaided stimulated Raman emission in neat 1, 2, dichloroethane with the addition of the dye DEST. This dye is weakly fluorescent, but shows strong ASE. We observed spectral narrowing of first Stokes line in DEST solutions. Further, we demonstrate pulse compression from 28 ps to 3 ps in the forward stimulated Raman emission. The single pass conversion efficiency is 20 % in 1 cm path length sample. Such enhanced efficiencies and compression offer tremendous promise for the design of tunable, ultrashort coherent light sources.

**Acknowledgements**

We thank Arvinder S. Sandhu for help with the experiments.

**Figure Captions**

Fig. 1 Experimental setup to measure the pulse duration

Fig. 2 Raman emission from 1, 2, Dichloroethane at input energy of 1.3 mJ

Fig. 3 Spectral evolution of Raman emission in $4 \times 10^{-5}$ mol/l DEST at input energies (a) 0.3 mJ (b) 0.7 mJ (c) 1.1 mJ (d) 1.3 mJ

Fig. 4 Dependence of Raman energy on the pump energy a) Neat solvent b) DEST solution

Fig. 5 Pulse duration of Raman emission (a) neat 1, 2, Dichloroethane (b) with $4 \times 10^{-5}$ mol/l DEST.

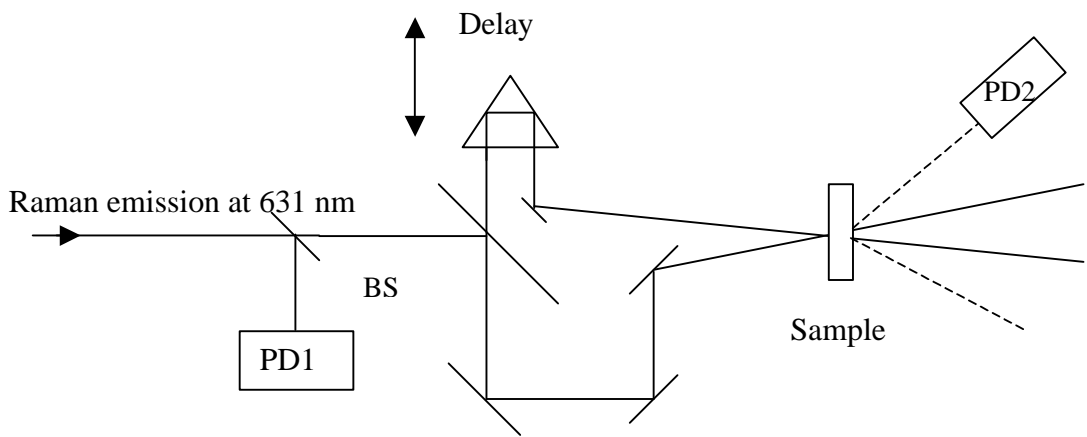

Fig. 1

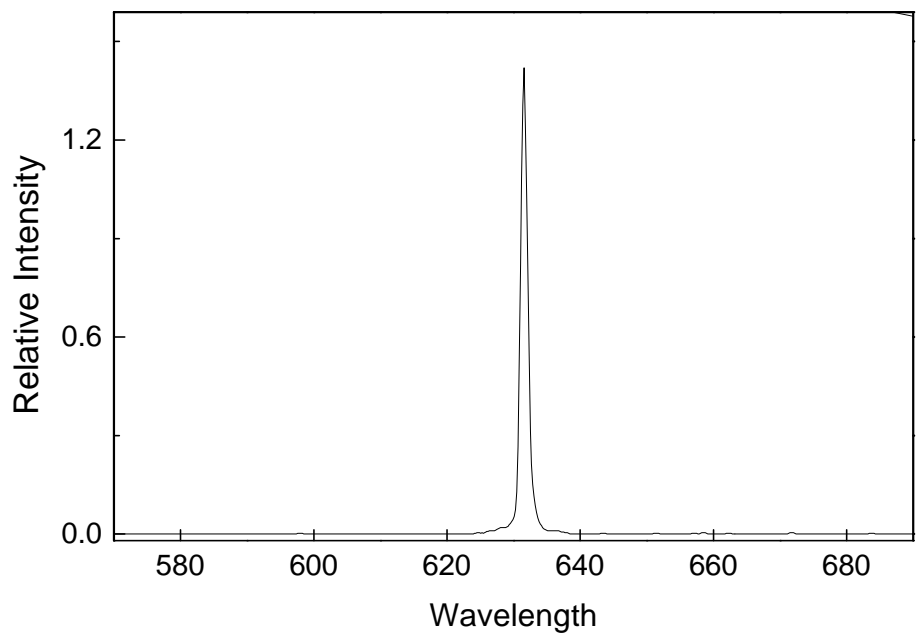

Fig.2

Dharmadhikari et.al.

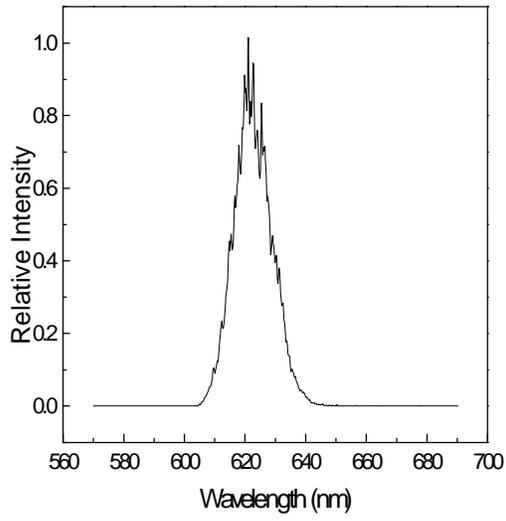

(a)

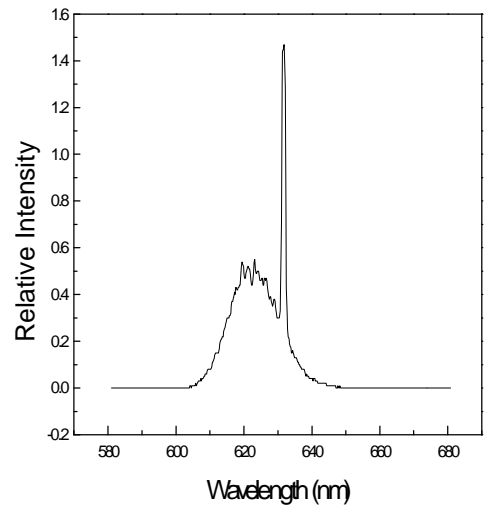

(c)

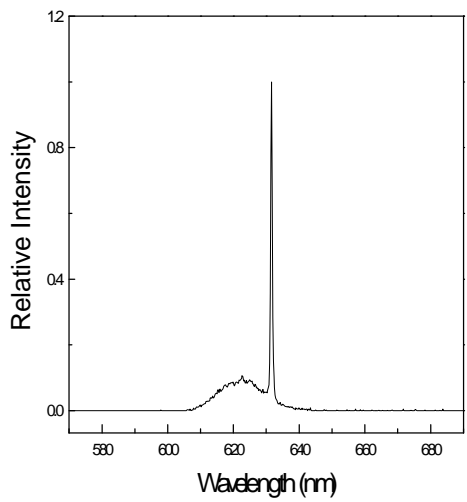

(c)

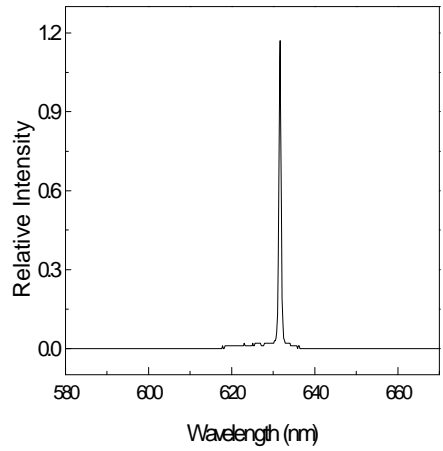

(d)

Fig. 3

Dharmadhikari et.al.

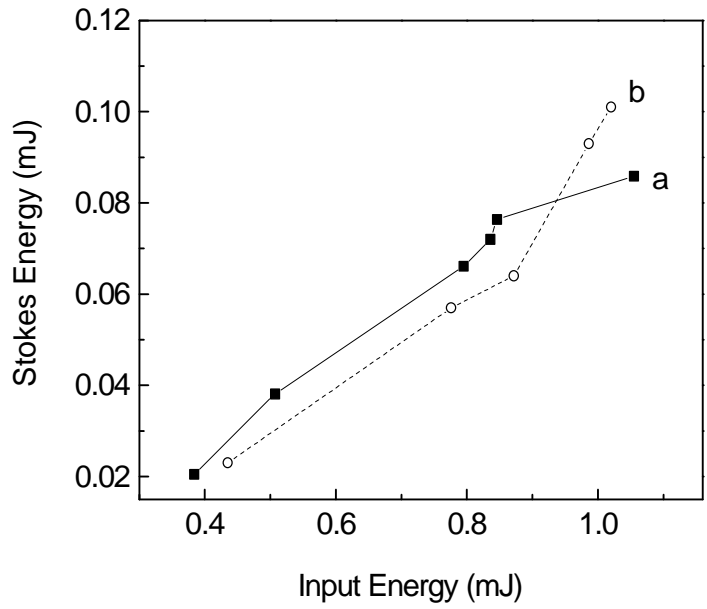

Fig. 4

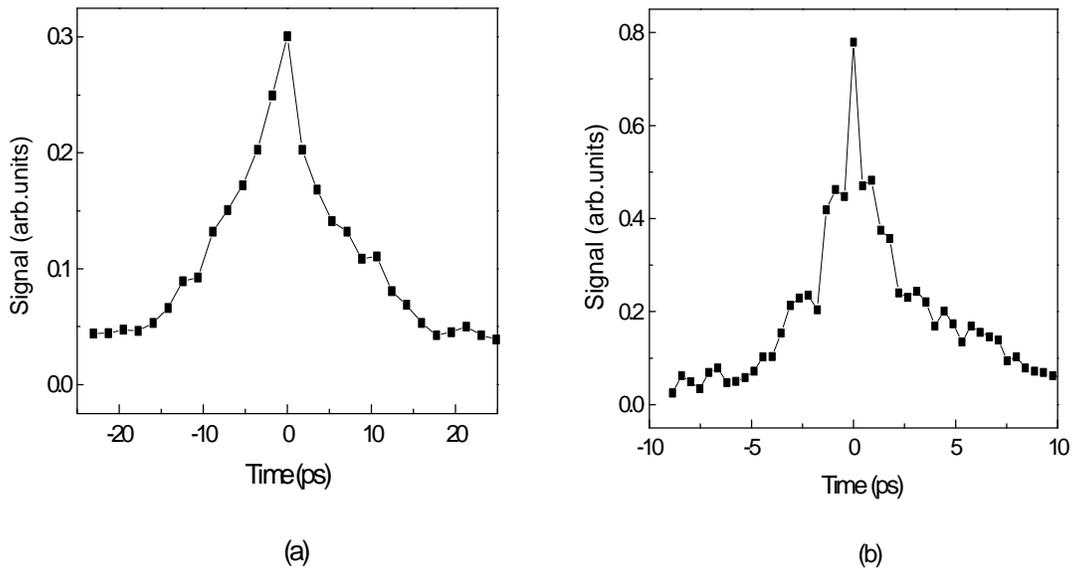

(a)                                  (b)

Fig. 5

Dharmadhikari et.al.